\documentclass[a4paper,11pt]{article}
\usepackage{pos}
\usepackage{physics}
\usepackage{wrapfig}
\usepackage{comment}
\def\bsxi{{\boldsymbol \xi}}
\def\bsx{{\boldsymbol x}}
\def\psibar{{\bar\psi}}

\renewcommand{\vec}[1]{\boldsymbol{#1}}
\newcommand{\be}{\begin{equation}}
\newcommand{\ee}{\end{equation}}
\newcommand{\bea}{\begin{eqnarray}}
\newcommand{\eea}{\end{eqnarray}}

\newcommand{\MSbar}{\overline{\mbox{MS}}}

\title{Non-perturbative thermal QCD at very high temperatures}
\ShortTitle{Non-perturbative thermal QCD at very high temperatures}

\author*[a,b]{Leonardo Giusti}
\author[a,b]{Matteo Bresciani}
\author[a,b]{Mattia Dalla Brida}
\author[d]{Tim Harris}
\author[c]{Davide Laudicina}
\author[b]{Michele Pepe}
\author[a,b]{Pietro Rescigno}

\affiliation[a]{Dipartimento di Fisica, Universit\`a di Milano-Bicocca, Piazza della Scienza 3, 
I-20126 Milano, Italy}

\affiliation[b]{INFN, Sezione di Milano-Bicocca, Piazza della Scienza 3, 
I-20126 Milano, Italy}

\affiliation[c]{Fakultät für Physik und Astronomie, Institut für Theoretische Physik II, Ruhr-Universität Bochum,\\
44780 Bochum, Germany}

\affiliation[d]{Institute for Theoretical Physics, ETH Zürich,
Wolfgang-Pauli-Str. 27, 8093 Zürich, Switzerland}

\emailAdd{leonardo.giusti@unimib.it}
\emailAdd{davide.laudicina@ruhr-uni-bochum.de}
\emailAdd{m.bresciani9@campus.unimib.it}
\emailAdd{mattia.dallabrida@unimib.it}
\emailAdd{harrist@phys.ethz.ch}
\emailAdd{michele.pepe@mib.infn.it}
\emailAdd{p.rescigno1@campus.unimib.it}

\abstract{We present a recently introduced strategy to study non-perturbatively thermal QCD up to temperatures
of the order of the electro-weak scale, combining step scaling techniques and shifted boundary conditions.
The former allow to renormalize the
theory for a range of scales which spans several orders of magnitude with a moderate
computational cost. Shifted boundary conditions avoid the need for the zero
temperature subtraction in the Equation of State. As a consequence, the simulated
lattices do not have to accommodate two very different scales, the pion mass and
the temperature. Effective field theory arguments guarantee that finite volume effects can be kept under
control safely. As a first application of this strategy, we present the results of the computation
of the hadronic screening spectrum in QCD with $N_f=3$ flavours of massless quarks for temperatures
from $T\sim 1$ GeV up to $\sim 160$ GeV.}

\FullConference{42nd International Conference on High Energy Physics (ICHEP2024)\\
18-24 July 2024\\
Prague, Czech Republic\\}

\begin{document}
\maketitle

\vspace{-0.75cm}

\section{Introduction}
\vspace{-0.375cm}

Quantum Chromodynamics (QCD) at very high temperatures plays a pivotal r\^ole in particle and nuclear physics as well as
in cosmology. In order to have a reliable and satisfactory understanding of the dynamics of the high temperature regime of QCD,
a fully non-perturbative approach is essential up to temperatures as high as the electro-weak
scale~\cite{PhysRevD.67.105008,Giusti:2016iqr,DallaBrida:2021ddx}. Here we present a strategy, first introduced for
the Yang-Mills theory in Ref. \cite{Giusti:2016iqr} and then generalized to QCD in Ref. \cite{DallaBrida:2021ddx}, which allows to simulate
the theory of strong interactions up to very high temperatures from first principles with a moderate computational effort.
As a first concrete implementation in QCD, we report the results that we have obtained for the
hadronic screening masses with $N_f=3$ massless quarks in a temperature interval ranging
from $T\sim 1$ GeV up to $\sim 160$ GeV~\cite{DallaBrida:2021ddx, Giusti:2024ohu}.
Those observables probe the exponential fall-off of two-point correlation functions of hadronic interpolating operators
in the spatial directions and are the inverses of spatial correlation
lengths, which characterize the response of the plasma when hadrons are injected into it. 
\vspace{-0.5cm}

\section{Non-perturbative thermal QCD at very high temperatures\label{sec:strategy}}
\vspace{-0.375cm}

\noindent {\it Renormalization and lines of constant physics $-$} A hadronic scheme is not a
convenient choice to renormalize QCD non-perturbatively when
considering a broad range of temperatures spanning several orders of magnitude. This
would require to accommodate on a single lattice both the temperature and the hadronic scale
which may differ by orders of magnitude, making the numerical computation extremely challenging. A similar problem
is encountered when renormalizing QCD non-perturbatively, and it was solved many years ago by introducing a
step-scaling technique~\cite{Luscher:1993gh}. 

In order to solve our problem, we have built on that knowledge by considering a non-perturbative definition
of the coupling constant, $\bar g^2_{\rm SF}(\mu)$, which can be computed precisely on the lattice for values
of the renormalization scale $\mu$ which span several orders of magnitude. Making a definite
choice, in this section we use the definition based on the Schr\"odinger functional (SF)~\cite{Luscher:1993gh}, however, notice that
other theoretically equivalent choices are available.
In particular, in our lattice setup we also made use of the gradient flow (GF) definition
of the running coupling ~\cite{Brida:2016flw,DallaBrida:2016kgh}, see appendix B of Ref. \cite{DallaBrida:2021ddx}.
Once $\bar g^2_{\rm SF}(\mu)$ is known in the continuum limit for $\mu \sim T$
\cite{Brida:2016flw,DallaBrida:2018rfy},
thermal QCD can
be renormalized by fixing the value of the running coupling constant at fixed lattice spacing $a$,
$g^2_{\rm SF}(g_0^2, a\mu)$,  to be
\vspace{-0.75cm}

\begin{align}
    g^2_{\rm SF}(g_0^2, a\mu) = \bar g^2_{\rm SF}(\mu)\; ,\qquad a\mu\ll 1\;.
\end{align}
This condition fixes the so-called lines of constant physics, i.e. the dependence of
the bare coupling constant $g_0^2$ on the lattice spacing, for values of $a$ at which the scale $\mu$ and
therefore the temperature $T$ can be easily accommodated. For a more complete discussion on how this technique is
implemented in practical lattice simulations we refer to appendix B of Ref. \cite{DallaBrida:2021ddx}.

\noindent {\it Shifted boundary conditions $-$} The thermal theory is defined by requiring that the fields satisfy shifted boundary
conditions in the compact direction~\cite{Giusti:2011kt,Giusti:2010bb,Giusti:2012yj},
while we set periodic boundary conditions in the spatial directions. The former consist
in shifting the fields by the spatial vector $L_0\, \bsxi$ when crossing the boundary in the
compact direction, with the fermions having in addition the usual sign flip. For the gauge
fields they read
\vspace{-0.5cm}

\begin{equation} \label{eq:shift_gluons}
  U_\mu(x_0+L_0,\bsx)= U_\mu(x_0,\bsx-L_0\bsxi)\; ,
  \quad
  U_\mu(x_0,\bsx+\hat{k}L_k)= U_\mu(x_0,\bsx)\; , 
\end{equation}
while those for the quark and the anti-quark fields are given by
\begin{align}
&\psi(x_0+L_0,\bsx)  =  -\psi(x_0,\bsx - L_0\bsxi)\; ,
\quad
& \psi(x_0,\bsx+\hat{k} L_k)  = \psi(x_0,\bsx)\; ,
\nonumber\\
&\psibar(x_0+L_0,\bsx)  =  -\psibar(x_0,\bsx - L_0\bsxi)\;,
\quad
&\psibar(x_0,\bsx+\hat{k} L_k)  = \psibar(x_0,\bsx)\; ,\label{eq:shift_quark}
\end{align}
\vspace{-0.75cm}

\noindent where $L_0$ and $L_k$ are the lattice extent in the 0 and $k$-directions respectively.
In the thermodynamic limit, a relativistic thermal field theory in the presence of a shift $\bsxi$ is equivalent to the 
very same theory with usual periodic (anti-periodic for fermions) boundary conditions but with a longer extension of the
compact direction by a factor $\sqrt{1+\vec \xi^2}$~\cite{Giusti:2012yj}, and thus the standard relation between
the length and the temperature is modified as $T=1/(L_0 \sqrt{1+\vec \xi^2})$. Shifted boundary conditions
represent a very efficient setup to tackle several problems that are otherwise very challenging both from the
theoretical and the numerical viewpoint. Some recent examples are provided by the SU($3$) Yang-Mills theory
Equation of State (EoS) which was obtained with a permille precision up
to very high temperatures~\cite{Giusti:2014ila,Giusti:2016iqr} and more recently in $N_f=3$ QCD with a novel
computation of the renormalization constant of the flavour-singlet local vector current \cite{Bresciani:2022lqc}. The same setup is
currently in use to carry out the first non-perturbative computation of the EoS at large temperatures in thermal
QCD \cite{DallaBrida:2020gux,Bresciani2024}. 

\noindent {\it Finite-volume effects $-$}
At asymptotically high temperatures, the mass gap developed by thermal QCD is proportional to $g^2 T$. On the other hand,
at intermediate temperatures, provided that the temperature is sufficiently large with respect to $\Lambda_{\rm QCD}$, the mass gap of the
theory is always expected to be proportional to the temperature times an appropriate power of the coupling constant.
As a consequence, when $LT\to \infty$ finite-size effects are exponentially suppressed in $LT$ times a
coefficient that decreases logarithmically
with the temperature. For this reason, we have always
employed large spatial extents, i.e. $L/a=288$, so that $LT$ ranges always from $20$ to $50$.

\noindent {\it Restricting to the zero-topological sector $-$} At high temperature, the topological charge distribution is expected to be highly peaked at zero. For QCD with three light degenerate flavours of mass $m$,
the dilute instanton gas approximation predicts for the the topological susceptibility  $\chi\propto m^3T^{-b}$ with $b\sim 8$.
The analogous prediction for the Yang--Mills theory has been verified explicitly on the lattice~\cite{Giusti:2018cmp}.
Similarly, computations performed
in QCD seem to confirm the $T$-dependence predicted by the semi-classical analysis even though the systematics due to the introduction of
dynamical fermions is still difficult to control ~\cite{Borsanyi:2016ksw}. As a result, already at low temperatures, namely at $T\sim1$ GeV,
the probability to encounter a configuration with non-zero topology in volumes large enough to keep finite volume effects under control is expected to
be several orders of magnitude smaller than the permille or so. For these reasons, we can safely restrict our calculations to the sector with zero topology.
\vspace{-0.375cm}

\section{Screening spectrum\label{sec:num}}
\vspace{-0.375cm}

As a concrete application of the strategy outlined in section \ref{sec:strategy}, we have performed numerical simulations at 12 values of the temperature,
$T_0, \ldots$, $T_{11}$ covering the range from approximately $1$~GeV up to about $160$~GeV. For the 9 highest ones, $T_0, \ldots$, $T_8$, gluons are
regularized with the Wilson plaquette action, while for the 3 lowest temperatures, $T_9$, $T_{10}$ and $T_{11}$, we adopt the tree-level
improved L\"uscher-Weisz gauge action. The three massless flavours are always discretized by the $O(a)$-improved Wilson--Dirac operator.
In order to extrapolate the results to the continuum limit, several lattice spacings are simulated at each temperature with the extension of the
fourth dimension being $L_0/a=4,6,8$ or $10$.
\begin{figure}
    \centering
    \includegraphics[width=0.4\textwidth]{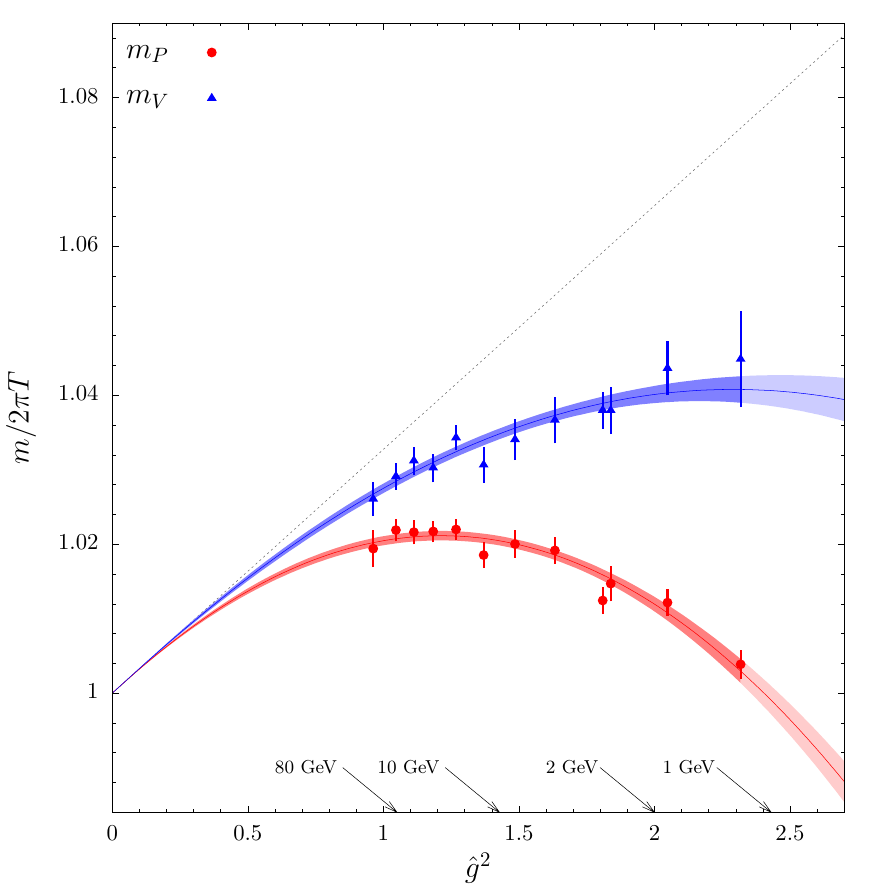} 
    \includegraphics[width=0.4\textwidth]{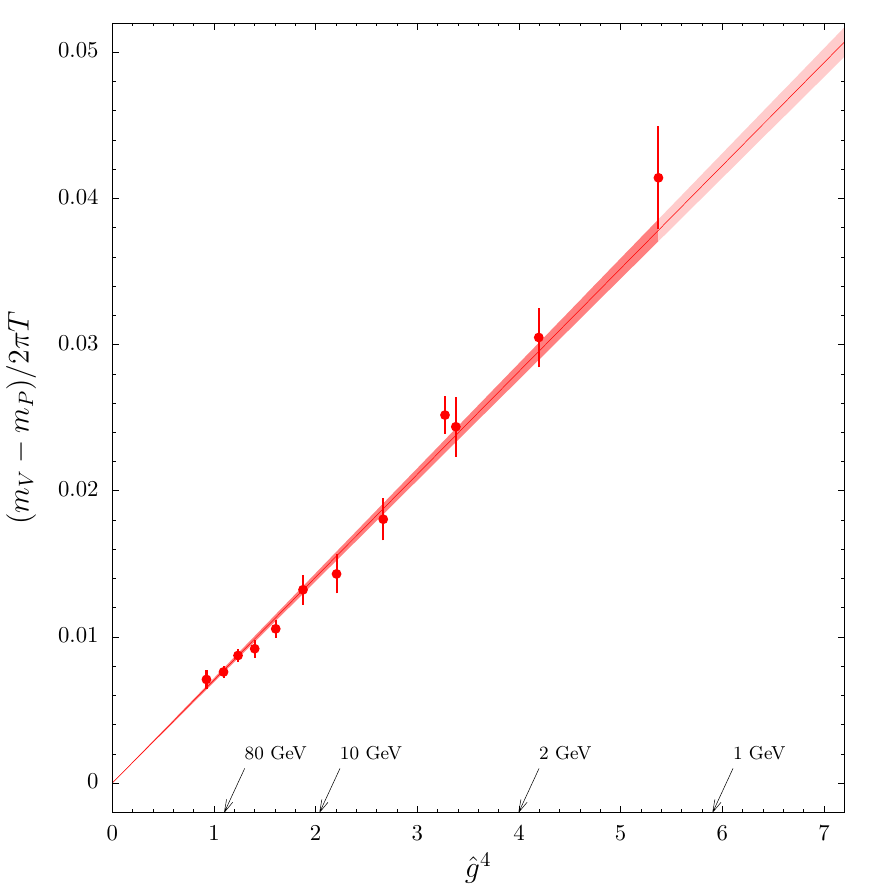} 
    \caption{Left: pseudoscalar (red) and vector (blue) screening masses versus $\hat g^2$. The bands represent the best fits in Eqs.~(\ref{eq:quartic}) and (\ref{eq:mvfit}), while the dashed line is the analytically known contribution.
    Right: the vector-pseudoscalar mass difference, normalized to $2\pi T$, versus $\hat g^4$. Red bands
    represent the best fits of the data as explained in the text.\label{fig:massCL}}
    \vspace{-0.5cm}
    
\end{figure}

\noindent {\it Mesonic screening masses $-$} The mesonic screening masses have been computed with a few permille accuracy in the continuum limit. Within our statistical precision, the screening masses
associated to the pseudoscalar and to the scalar density are found to be degenerate and a similar discussion holds
for the screening masses of the vector and of the axial current as expected in the presence of chiral symmetry restoration.
Given the high accuracy of our non-perturbative data, it has been possible to parameterize the temperature dependence of
the masses. In order to do that, we have introduced the function $\hat g^2 (T)$ defined as 
\begin{equation}\label{eq:gmu}
  \frac{1}{\hat g^2(T)} \equiv \frac{9}{8\pi^2} \ln  \frac{2\pi T}{\Lambda_{\MSbar}} 
  + \frac{4}{9 \pi^2} \ln \left( 2 \ln  \frac{2 \pi T}{\Lambda_{\MSbar}}  \right)\; , 
\end{equation}
where $\Lambda_{\MSbar} = 341$~MeV is taken from Ref.~\cite{Bruno:2017gxd}. It corresponds
to the 2-loop definition of the strong coupling constant in the $\MSbar$ scheme at the renormalization scale
$\mu=2\pi T$. For our purposes, however, this is just a function of the temperature $T$
that we use to analyze our results and which makes it easier to compare with the known perturbative results.

The temperature dependence of the pseudoscalar mass has been parameterized with a quartic polynomial in $\hat{g}$ of the form
\vspace{-0.75cm}

\begin{align}
\label{eq:quartic}
    \frac{m_P}{2\pi T} = p_0 + p_2 \hat{g}^2 + p_3 \hat{g}^3 + p_4 \hat{g}^4 \, .
\end{align}
The leading and the quadratic coefficients have been found to be compatible with the free theory value and the next-to-leading order
correction, i.e. $p_0=1$ and $p_2=0.0327$ \cite{Laine:2003bd}. Once $p_0$ and $p_2$ have been fixed to their
corresponding perturbative values, we obtain for the
cubic and the quartic fit parameters $p_3=0.0038(22)$ and $p_4=-0.0161(17)$
with ${\rm cov}(p_3,p_4)/[\sigma(p_3)\sigma(p_4)]=-1.0$ with an
excellent $\chi^2/{\rm dof} =0.75$. Such a polynomial is displayed, as a red band, together with the non-perturbative data in the left panel of
Figure~\ref{fig:massCL}. It is clear that the quartic term is necessary to explain the behaviour of the non-perturbative data in the entire
temperature range. In particular, at the highest temperatures it contributes for about $50\%$ of the total contribution due to the 
interactions, while at low temperature it competes with the quadratic coefficient to bend down $m_P/2\pi T$.

The mass difference between the vector and the pseudoscalar mass is due to spin-dependent contributions which
are of $O({g}^4)$ in the effective field theory. By plotting our results versus $\hat{g}^4$,
see right panel of Figure~\ref{fig:massCL}, these turn out to lie on a straight line with vanishing intercept in the entire range of temperature.
We then parameterized the temperature dependence with
\vspace{-0.5cm}

\begin{align}
    \label{eq:spindep}
    \frac{(m_{V} - m_{P})}{2\pi T} = s_4\, \hat g^4\;  
\end{align}
and we obtain $s_4=0.00704(14)$ with $\chi^2/{\rm dof}=0.79$. It is remarkable that, even at the highest temperatures which was simulated, the mass
difference is clearly different from zero, a fact which is not expected by the next-to-leading order estimate
obtained in the effective field theory. The best parameterization for the vector screening mass is thus given by
\vspace{-0.5cm}

\begin{align}
    \label{eq:mvfit}
    \frac{m_{V}}{2\pi T} = p_0 + p_2\, \hat g^2 + p_3\, \hat g^3 + (p_4 + s_4) \, \hat g^4\; ,
\end{align}
with covariances ${\rm cov}(p_3,s_4)/[\sigma(p_3)\sigma(p_4)]=0.08$ and ${\rm cov}(p_4,s_4)/[\sigma(p_4)\sigma(p_4)]=-0.07$. In the vector channel the
quartic contribution appearing in Eq.~\eqref{eq:mvfit} is responsible for about $15\%$ of the total contribution due to interaction at the
electro-weak scale.
Moreover, the quartic coefficient for the vector mass is about $50\%$ smaller than the corresponding coefficient for
the pseudoscalar channel.
\begin{wrapfigure}{r}{0.5\textwidth}
\centering
    \includegraphics[width=\linewidth]{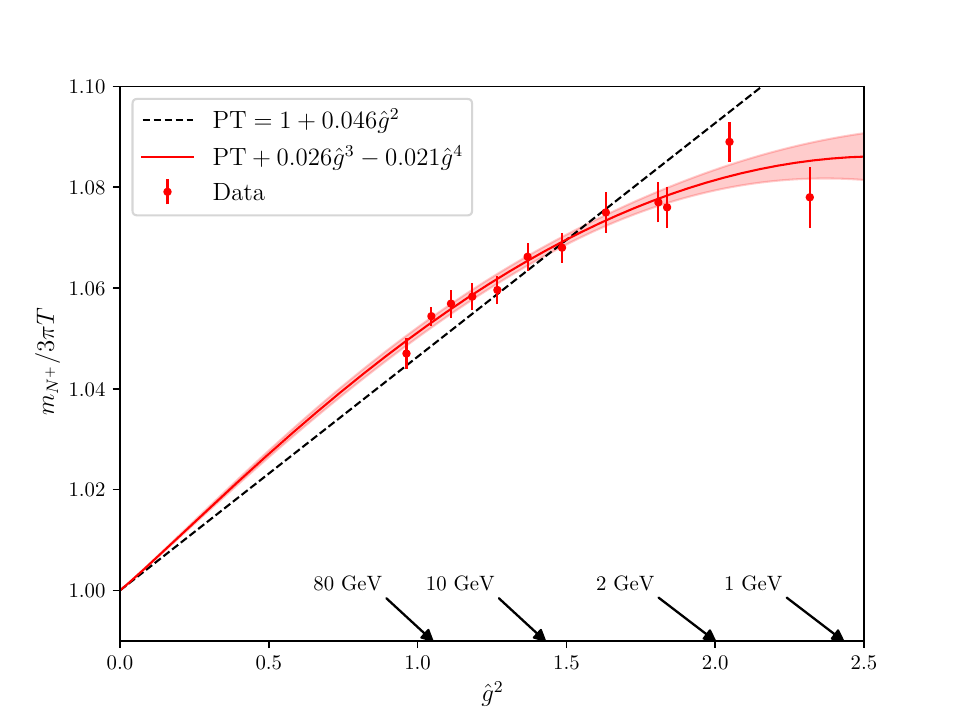}
    \caption{Nucleon screening mass versus $\hat g^2$. The band      represent the best fit to Eq.~(\ref{eq:quartic-baryon}), while the dashed line
      is the analytically known contribution.\label{fig:mass-baryon-CL}}
\end{wrapfigure}
As a consequence, its contribution is not large enough to compete with the quadratic coefficient and to bend down the
value of the vector mass at low temperature. For a detailed analysis of the results see section 7 of Ref. \cite{DallaBrida:2021ddx}.\\
\noindent {\it Baryonic screening masses $-$}
In contrast with the mesonic case, there are very few studies on the baryonic sector both on the lattice
and in the three dimensional effective theory and, for what concerns lattice calculations, no continuum limit extrapolation
has ever been performed. In Ref.~\cite{Giusti:2024ohu} we have computed the baryonic screening masses for the first time in the
continuum limit and with a final accuracy
of a few permille from $1$ GeV up to the electro-weak scale. As expected in a chirally symmetric regime, the positive and the negative parity screening
masses are found to be degenerate in the entire range of temperatures. For this reason, in the following we only focus on the positive parity
mass $m_{N^+}$. The final results are shown in Figure~\ref{fig:mass-baryon-CL} as a function of $\hat{g}^2(T)$. 

As it is clear from the plot, the bulk of the baryonic screening mass is given by the free field theory $3\pi T$ plus a $4-8\%$ positive contribution
due to interaction. It is rather clear that from $T\sim160$ GeV down to $T\sim5$ GeV the perturbative expression is within half a percent with respect to
the non-perturbative data. The full set of data, however, shows a distinct negative curvature which requires higher orders in the coupling constant to be
parameterized. Similarly to the case of the mesonic screening masses, the temperature dependence of the baryonic screening mass has been parameterized
with the ansatz
\vspace{-0.75cm}

\begin{align}
    \label{eq:quartic-baryon}
\frac{m_{N^+}}{3\pi T} = b_0 + b_2\, \hat g^2 + b_3\, \hat g^3 + b_4\, \hat g^4\; .  
\end{align}
The coefficients $b_0$ and $b_2$ turn out to be compatible with the free-theory and the next-to-leading values,
i.e. $b_0=1$ and $b_2=0.046$ \cite{Giusti:2024mwq}. Then, by
enforcing those values and fitting again, we obtain $b_3=0.026(4)$, $b_4=-0.021(3)$ and
${\rm cov}(b_3,b_4)/[\sigma(b_3) \sigma(b_4)]=-0.99$ with $\chi^2/{\rm dof}= 0.64$, which is the best parameterization of our results over the entire range of
temperatures explored. Notice that, in general, other parameterizations of the lattice data are possible as well. These, however, result in the disagreement
between the fit result for $b_2$ and the 1-loop perturbative correction. For a more detailed discussion on such parameterizations
we refer to section 5 of Ref. \cite{Giusti:2024ohu}.

We conclude by noticing that the evaluation of the EoS in a similar temperature range, computed with the strategy
outlined in Ref.~\cite{Bresciani2024}, is almost completed. Preliminary results shown at the conference are not
reported here for lack of space.
\vspace{-0.5cm}

\bibliographystyle{JHEP}
\bibliography{bibfile}

%\begin{thebibliography}{99}
%\bibitem{...}
%....

%\end{thebibliography}

\end{document}